\documentclass[sigconf]{acmart}
\usepackage{cleveref}
\usepackage{amsmath}
\usepackage{enumitem}
\usepackage{multirow}
\usepackage{comment}
\usepackage{caption}
\usepackage{titlesec}
\titlespacing*{\section}{1pt}{0.25\baselineskip}{0.25\baselineskip}
\titlespacing*{\subsection}{1pt}{0.25\baselineskip}{0.25\baselineskip}

\newcommand{\squishlist}{
 \begin{list}{$\bullet$}
  { \setlength{\itemsep}{0pt}
     \setlength{\parsep}{2pt}
     \setlength{\topsep}{2pt}
     \setlength{\partopsep}{0pt}
     \setlength{\leftmargin}{1.5em}
     \setlength{\labelwidth}{1em}
     \setlength{\labelsep}{0.5em} } }

\newcommand{\squishend}{
\end{list}  }

\DeclareMathOperator*{\argmax}{arg\,max}

\AtBeginDocument{%
  }

\begin{document}

\author{Allen Lin*, Jianling Wang, James Caverlee}
\affiliation{
\institution{Texas A\&M University}
\city{College Station}
\state{TX}
\country{USA}}
\email{al001@tamu.edu, jlwang@tamu.edu, caverlee@tamu.edu}

\author{Ziwei Zhu}
\affiliation{%
  \institution{George Mason University}
  \city{Fairfax, Virginia}
  \country{USA}
}
\email{zzhu20@gmu.edu}

\title{Federated Conversational Recommender System}

\begin{abstract}
Conversational Recommender Systems (CRSs) have become increasingly popular as a powerful tool for providing personalized recommendation experiences. By directly engaging with users in a conversational manner to learn their current and fine-grained preferences, a CRS can quickly derive recommendations that are relevant and justifiable. However, existing conversational recommendation systems (CRSs) typically rely on a centralized training and deployment process, which involves collecting and storing \textit{explicitly-communicated user preferences} in a centralized repository. These fine-grained user preferences are completely \textit{human-interpretable} and can easily be used to infer sensitive information (e.g., financial status, political stands, and health information) about the user, if leaked or breached. To address the user privacy concerns in CRS, we first define a set of privacy protection guidelines for preserving user privacy under the conversational recommendation setting. Based on these guidelines, we propose a novel federated conversational recommendation framework that effectively reduces the risk of exposing user privacy by (i) de-centralizing both the \textit{historical interests estimation} stage and the \textit{interactive preference elicitation} stage and (ii) strictly bounding privacy leakage by enforcing \textit{user-level differential privacy} with meticulously selected privacy budgets. Through extensive experiments, we show that the proposed framework not only satisfies these user privacy protection guidelines, but also enables the system to achieve competitive recommendation performance even when compared to the state-of-the-art non-private conversational recommendation approach.
\end{abstract}

\maketitle

\section{Introduction}
\label{sec:intro}
Conversational recommender systems (CRSs) have gained significantly increasing attention in the research community for their wide range of applications and promising functionality, spanning from streaming services \cite{CRIF} to E-Commerce \cite{UNICORN}. These systems rely heavily on the elicited user preferences to make personalized recommendations, which are typically conveyed in a human-interpretable manner and could contain detailed description of users' personal needs and interests. For example, a user might come to the CRS asking for an \textit{affordable} restaurant that only serves \textit{vegetarian} food in \textit{Austin, Texas}. While these communicated preferences greatly assist the CRS in making personalized recommendations, such information, if leaked or breached, could easily be exploited to infer sensitive information (e.g., financial status, dietary restriction, and location) about the user \cite{massa2009trust, minto2021stronger_rec21}. Despite the success of interactive preference elicitation, \textit{addressing user privacy concerns in CRSs} remains an unexplored research area.

Current CRSs \cite{crm, ear, scpr, UNICORN, CRIF} are trained and deployed in a \textit{centralized} manner, meaning that the communicated, human-interpretable, preferences information of all users is \textit{fully accessible} to the service provider. Such a centralized recommendation framework raises serious concerns of user privacy leakage or unintended data breaches \cite{gao_sigir_2020dplcf}. While many strategies \cite{jin2022federated, yang2020federated_rec, ammad2019federated_cf} have been proposed to address the privacy concerns in a \textit{static} recommendation setting, these approaches mainly focus on privatizing users' raw item-interaction history (e.g., in the form of a real-valued matrix), which is not directly applicable to protecting the user preferences that are communicated during the \textit{interactive preference elicitation} process in a conversational recommendation setting.

Recognizing the significant research gap, we first define a set of guidelines for preserving user privacy in a conversational recommendation setting: 
(i) \textbf{User Information Locality}: the explicitly communicated preferences and the interaction history (personal feedback) of all users should never be uploaded to the server; instead, they should only be stored locally at each user's own devices; 
(ii) \textbf{User Contribution Anonymity}: while users never upload any personal data (e.g., interaction history or stated preferences) to the server, they are allowed to contribute to the tuning of the CRS. Such contribution (e.g., via parameter gradients), however, should always be protected with the \textit{local differential privacy} guarantee -- to prevent the risk of reverse engineering\footnote{E.g., uploaded model gradients can be used to infer the actual user behaviors.} \cite{gao_sigir_2020dplcf}; and  
(iii) \textbf{Interaction and Recommendation Locality}: all interactions (e.g., conversations) between the system and the user should only be stored locally at the user's own devices. And all recommendations should also be inferred locally at a user's own devices. 

Following the aforementioned privacy protection guidelines, we propose a novel federated conversational recommendation framework -- FedCRS. Specifically, it starts with building a predictive model to form an initial estimation of users' historical interests by learning a set of latent representations from the historical user-item interactions. To satisfy \textbf{user information locality}, instead of uploading their personal interaction history to the server, users only upload gradient updates needed to tune the predictive model. Before uploading, these gradient updates are perturbed with a privatization mechanism to ensure user-level local differential privacy, satisfying \textbf{user contribution anonymity}. After obtaining the initial estimation of users' historical interests, a reinforcement learning based policy agent is deployed to initiate a personalized interactive preference elicitation process with each user. To account for the \textit{environment heterogeneity\footnote{A commonly encountered issue in federated reinforcement learning that hinders a uniform policy to deliver optimal interaction experience to every user \cite{jin2022federated}. E.g., for movie recommendation, one user might care more about the \textit{genre} while another cares more about the \textit{director}.}} caused by each user having unique needs and preferences, a local user embedding projection layer is integrated with the policy agent to finetune the uniform interaction policy to further personalize each user's recommendation experience. This policy agent is trained in a similar fashion as the predictive model with each user uploading the gradient updates instead of the actual stated personal preferences. Upon completion of the tuning process, a copy of the policy agent is sent to every user to be used to infer recommendations locally, satisfying \textbf{interaction and recommendation locality}. In summary, the main contributions of this work are as follows:
\squishlist
\item To our best knowledge, this is the first work to comprehensively study user privacy concerns under the conversational recommendation setting and introduce a set of guidelines to be satisfied during training and inference to protect user privacy.  
\item We propose the FedCRS framework with unique design and privatization mechanisms that provides personalized recommendation experiences to users while removing the need of centralized collection of any personal data. 
\item We show that the proposed FedCRS framework is able to achieve competitive recommendation performance when compared to the state-of-the-art conversational recommendation approach, while providing strong user privacy protection.
\squishend

\section{Preliminaries}
\subsection{Conversational Recommender System}
\label{prelim_crs}
A CRS generally starts with a \textit{historical interests estimation} stage where it utilizes a user's collected personal feedback (item-interaction history) -- which can be either explicit (e.g., movie ratings) or implicit (e.g., views or clicks) -- to establish an initial estimation of each user. This is typically done via building a predictive model (e.g., a factorization machine \cite{FM} or a graph neural network \cite{gnn}) to learn a set of embeddings that encodes the estimated historical interests of users based on their historical item-interactions. 

However, as introduced in \cite{jannach2021survey, survey2, survey3}, historical interactions can often be noisy and might fail to represent a user's current interests, as preferences might have changed over time. Thus, to further refine the initial estimation, the system then initiates a personalized \textit{interactive preference elicitation} stage, in which the system directly engages with the user in a conversational manner to inquire about the user's current and fine-grained preferences. In many recent CRSs \cite{crm, ear, scpr, UNICORN, FPAN, CRIF}, this personalized preference elicitation stage is governed by a reinforcement learning based policy agent that decides how to interact with the user. Formally, let \begin{math} V \end{math} denote the itemset, and \begin{math} P = (p_0, p_1,..., p_m) \end{math} denote a set of $m$ domain-specific attributes (e.g., movie genres) that describe an item \begin{math} v \in V \end{math}. Based on the system's current understanding of the user, the policy agent decides whether to ask the user more attributes or make a recommendation. If the policy agent thinks not enough preference evidence has been collected, it will pick one or more attribute(s) \begin{math} p \end{math} from the set of unasked attributes \begin{math} P_{cand} \end{math} to prompt the user. If the user confirms the asked attribute, then policy agent updates the collected (attributed-based) user preferences, \begin{math} P_{u} \end{math}, by adding \begin{math} p \end{math} to \begin{math} P_u \end{math}; otherwise \begin{math} P_u  \end{math} remains unchanged. If the policy agent decides enough information has been collected, the CRS then ranks all the candidate items and recommends the top k ranked items to the user. If the user accepts the recommendation, then the system quits. If the user rejects all the recommended items, this cycle continues until the user quits or reaching a predefined number of turns.

\subsection{Federated Learning}
Federated Learning \cite{federated_survey, federated_survey2, federated_survey3} is a privacy-preserving technique that leverages the rich private data of users to collaboratively train a central model without the need to collect them. It was originally proposed as a way to train a central model on privacy-sensitive data distributed across users’ devices. In the federated learning paradigm, the user’s data never has to leave the client. Instead, clients train a local model on their private data and share model updates with the server. These updates are then aggregated before a global model update on the central model is performed. Finally, a copy of updated central model is sent back to each client and the process is repeated until convergence or satisfying performance has been achieved. Since model updates usually contain much less information than the raw user data, the risks of privacy leakage is effectively reduced \cite{chai2020secure, yang2020federated_rec, ammad2019federated_cf}. In this work, we develop a Federated Learning framework for training and deploying conversational recommender systems in a distributed manner to remove the need for users to upload neither their explicitly stated preferences nor their past interaction history, better protecting users' privacy. 

\subsection{Local Differential Privacy}
\label{subsect:LDP}
Differential privacy \cite{dwork2006differential, dwork2008differential} has become the gold standard for strong privacy protection. It provides a formal guarantee that a model’s results are negligibly affected by the participation of any individual client \cite{lee2011much}. Differential privacy was initially introduced in a centralized setting where a trusted aggregator collects raw user data and enforces some privatization mechanism (e.g., injecting controlled noise) either in the query inputs, outputs, or both. However, the assumption of that the aggregator can be trusted can often be unrealistic \cite{arachchige2019local, cormode2018privacy}. Thus, the notion of \textit{local differential privacy} is proposed. Instead of injecting noise at the aggregator, the noise injection is done by the client before uploading. Since the aggregator is no longer assumed to be trusted, the local version of differential privacy provides much stronger user privacy protection \cite{wang2020comprehensive}. 

Formally, let $U$ be a set of clients, and each $u$ has a private value $v$ to be sent to an untrusted aggregator. Local differential privacy can guarantee that the leakage of private information for each user is strictly bounded via applying a randomized algorithm $M$ to the private value $v$ and sending the perturbed value $M(v)$ to the aggregator. In order for the privacy guarantee to hold, the randomized algorithm $M$ needs to satisfy $\epsilon$-local differential privacy which is true if and only if the following inequality holds \cite{minto2021stronger_rec21, qi2020privacy_news}:
\begin{equation*}
    Pr[M(v) = y] \leq e^{\epsilon} Pr[M(v') = y], 
\end{equation*}
where $v$ and $v'$ are two arbitrary input private values, and $y \in range(M)$. The $\epsilon$, ($\geq 0$), is called the privacy budget, where it controls the trade-off between utility of the perturbed data and the degree of privacy preserved. When $\epsilon = 0$, we achieve perfect privacy preservation but zero utility, while for $\epsilon = \infty$ we would achieve zero privacy preservation but perfect utility. In this paper, we use the local differential privacy technique in conjunction with our federated learning framework to train a CRS. Specifically, we apply a randomized algorithm $M$ to privatize the model gradient updates uploaded to the server, significantly reducing the risk of privacy leakage through reverse engineering \cite{gao_sigir_2020dplcf}.

\section{Proposed Framework}
In this section, we introduce a novel \textbf{Fed}erated learning framework for preserving user privacy in \textbf{C}onversational \textbf{R}ecommender \textbf{S}ystems -- FedCRS. Specifically, it is designed to tackle three key challenges: (i) how to privatize the historical user-item interactions during the \textit{historical user interests estimation} stage; (ii) how to privatize the explicitly communicated user preferences information during the \textit{interactive preference elicitation} stage; and (iii) how to privatize the interactive conversational recommendation process. 

FedCRS starts with the \textit{decentralized historical interests estimation} stage. To prevent user information leakage, we integrate federated learning and local deferential privacy mechanisms to design a decentralized training paradigm. Concretely, each user's learned embedding is kept locally and the gradient updates are perturbed to guarantee user-level differential privacy before uploading - satisfying both \textbf{user information locality} and \textbf{user contribution anonymity}. Upon obtaining the estimated historical interests, FedCRS enters the \textit{decentralized interactive preference elicitation} stage. In this phase, a policy agent is deployed to learn a personalized interaction policy, instructing the CRS on how to engage with users to gain a more comprehensive understanding of their needs and preferences. However, to deliver a personalized preference elicitation experience, the policy agent must first be trained through a trial-and-error manner that involves interacting and receiving feedback from users. Thus, to privatize each user's communicated preferences when interacting with the policy agent, a similar decentralized training paradigm is adopted, such that the actual communicated preferences information of each user are kept locally and only the perturbed policy gradient updates are uploaded to the server. In addition, since one uniform interaction policy fails to provide optimal elicitation experience for all users \cite{jin2022federated}, a local user embedding projection layer is integrated and trained in conjunction with the policy agent, to further personalize the learned policy.    

Once the training has been completed, a copy of the predictive model and the policy agent is distributed to each user. This allows the CRS to interact with the user and generate recommendations on each user's own devices -- satisfying \textbf{interaction and recommendation locality}. Note, while there are works that decentralize the training and deployment of recommender systems, they either do not guarantee user-level local differential privacy \cite{rw1_no_ldp, rw2_no_ldp, rw3_no_ldp}, failing to satisfy \textbf{user contribution anonymity}, or cannot easily generalize to a conversational recommendation setting \cite{rw4_no_crs, rw5_no_crs}, failing to satisfy \textbf{interaction and recommendation locality}.

\begin{figure}
  \includegraphics[width=.5\textwidth]{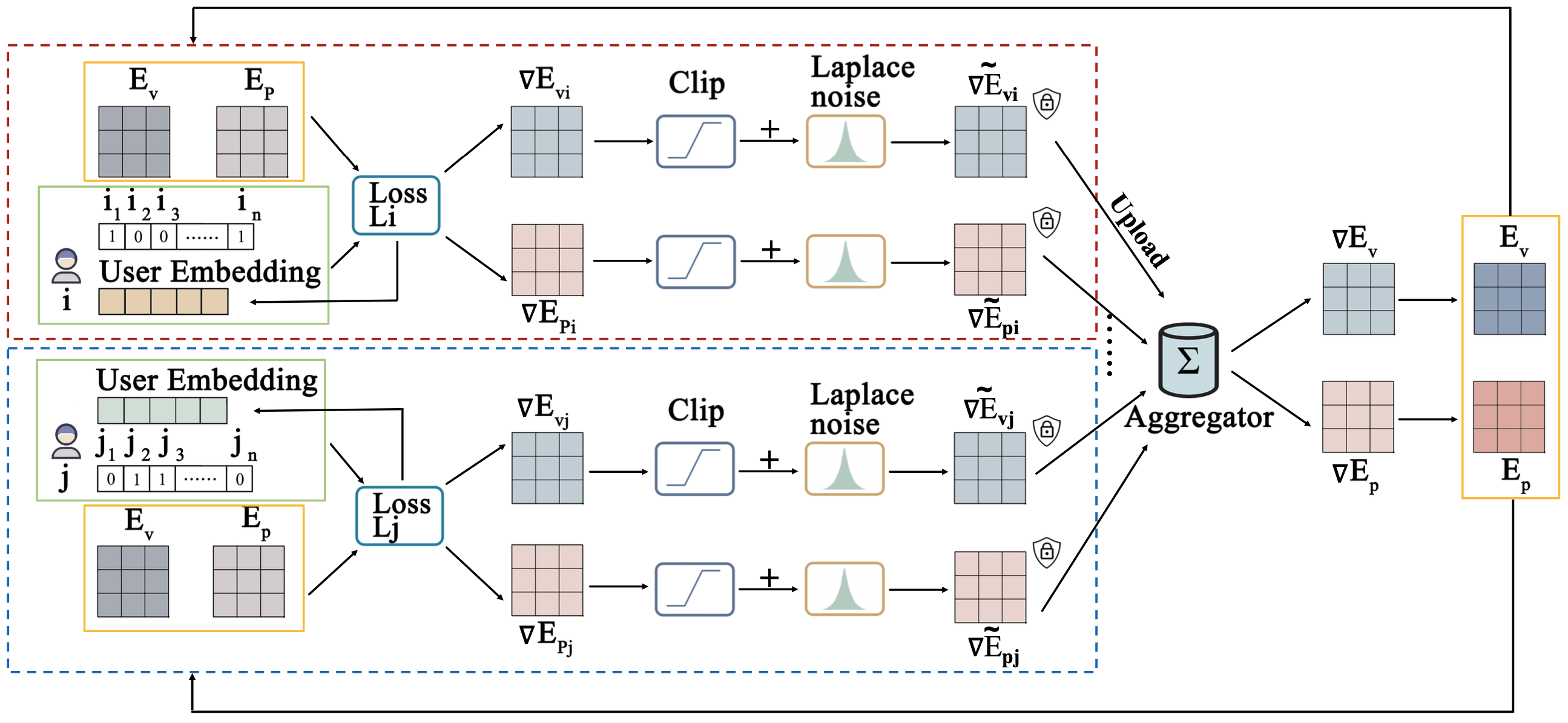}
  \caption{The overall flow of the decentralized historical interests estimation stage}
  \Description{}
  \label{fig:model1}
\end{figure}

\subsection{Decentralized Historical Interests Estimation}
\label{stage1}
For a CRS to provide a personalized recommendation experience, the system must first establish an initial understanding of the user based on his or her historical item-interactions. Thus, the goal of the \textit{Decentralized Historical Interests Estimation} stage is to build a predictive model to learn a set of latent embeddings for all users, items, and attributes – denoted respectively as – $\textit{\textbf{E}}_U$, $\textit{\textbf{E}}_V$, $\textit{\textbf{E}}_P$ -- in a decentralized way to protect users' privacy. $\textit{\textbf{E}}_U$ denotes the set of user embeddings that encodes the estimated historical interests of each user, while $\textit{\textbf{E}}_V$ and $\textit{\textbf{E}}_P$ represent two sets of embeddings that respectively encode the latent characteristics or properties of each item or attribute. Following \cite{FPAN, scpr, ear}, we choose the factorization machine (FM) \cite{FM} as the predictive model\footnote{Note that the choice of predictive model is flexible; we choose FM due to its demonstrated success in CRSs}. Given user $u$, the user's explicitly stated preferences $P_u$, and an item $v$, $u$'s predicted interest in $v$ is computed as:
\begin{equation*}
    y(u, v, P_u) = \textbf{e}_u^T\textbf{e}_v + \sum_{p_i \in P_u} \textbf{e}_v^T\textbf{e}_{p_i}
\end{equation*}
where $\textbf{e}_u$ and $\textbf{e}_v$ respectively denote the latent embeddings of user $u$ and item $v$, and $\textbf{e}_{p_i}$ denotes the latent embeddings for attribute $p_i \in P_u$. To train the predictive model, we adopt the dual-loss pairwise Bayesian Personalized Ranking (BPR) objective function:
\begin{equation}
 .  \begin{split}
    \label{eq:dual_loss_bpr}
    L = \sum_{(u,v,v') \in D_1} -ln\sigma(y(u,v,P_u) - y(u,v',P_u)) \ +  
    \\ \sum_{(u,v,v') \in D_2} -ln\sigma(y(u,v,P_u) - y(u,v',P_u)) + \lambda_{\theta}||\theta||^2
    \end{split}
\end{equation}
where $D_1$ denotes the set of pairwise instances where each pair is composed of an interacted item and a randomly chosen uninteracted item - same as in the traditional BPR setting. $D_2$, on the other hand, consists of instance pairs such that both the interacted and the uninteracted items have attributes matching $P_u$. $\sigma$ denotes the sigmoid function and $\lambda_{\theta}$ denotes the regularization parameter to prevent overfitting. Note, as shown in many previous works \cite{ear, scpr, FPAN}, including the second part of the loss function is crucial for training the model to correctly rank among the candidate items. 

In order to train the predictive model, the majority of existing CRSs \cite{crm, ear, scpr, UNICORN, FPAN, CRIF} employ a centralized paradigm \cite{massa2009trust}, which necessitates users to share their raw personal item-interaction history with a central server. This approach, however, overtly breaches the \textbf{user information locality} guideline and poses a significant risk of privacy leakage. To address this concern, we introduce a decentralized training paradigm that integrates federated learning and local differential privacy protection mechanisms -- as illustrated in \Cref{fig:model1}. Within the proposed paradigm, item-interaction data remain securely on users' devices, while the predictive model is collaboratively trained through the aggregated contributions of multiple different users.

Initially, each client\footnote{Following \cite{qi2020privacy_news}, we regard each user device that participates in the training of the model as a \textit{client}.} randomly initializes an F-dimensional local user embedding $\textit{\textbf{e}}_u$ which represents the estimated historical interests of the user $u$. This user embedding never leaves the client. Then, each client receives a client-agnostic item matrix $\textit{\textbf{E}}_V$ and a client-agnostic attribute matrix $\textit{\textbf{E}}_P$ from the server, signalling the start of a federated epoch. Based on the accumulated item-interaction history, the local user embedding $\textit{\textbf{e}}_u$, and the received $\textit{\textbf{E}}_V$ and $\textit{\textbf{E}}_P$, each client, with respect to \Cref{eq:dual_loss_bpr}, computes (i) a local gradient update $\nabla\textit{\textbf{e}}_u$ for the local user embedding $\textit{\textbf{e}}_u$, which is, again, kept strictly locally; (2) a local gradient update $\nabla\textit{\textbf{E}}_{V_u}$ for the item matrix $\textit{\textbf{E}}_V$; and (3) a local gradient update $\nabla\textit{\textbf{E}}_{P_u}$ for the attribute matrix  $\textit{\textbf{E}}_P$. 

The computed local user embedding gradient update $\nabla\textit{\textbf{e}}_u$ is directly used to \textit{locally} update the current local user embedding $\textit{\textbf{e}}_u$ as: 
\begin{equation}
\label{u_emb_update}
    \textit{\textbf{e}}_u = \textit{\textbf{e}}_u - \eta_1 \cdot \nabla \textit{\textbf{e}}_u,
\end{equation}
where $\eta_1$ is the learning rate used to optimize the $\textit{\textbf{e}}_u$. Meanwhile, the computed local gradient update for the item and attribute matrices -- $\nabla\textit{\textbf{E}}_{V_u}$ and $\nabla\textit{\textbf{E}}_{P_u}$ -- are uploaded and aggregated on the server. Although, the local gradient updates, $\nabla\textit{\textbf{E}}_{V_u}$ and $\nabla\textit{\textbf{E}}_{P_u}$, are considered safer to upload than the user's raw item-interaction history, they may still be reverse-engineered to reveal sensitive personal information about the user \cite{gao_sigir_2020dplcf}. Hence, to offer a more rigorous user privacy protection, we further apply a client-level local differential privacy mechanism to the local gradients with the randomized algorithm, $M$, defined as follows: 
\begin{equation}
\label{stage1_lap}
    M(\nabla\textit{\textbf{E}}) = clip(\nabla\textit{\textbf{E}}, \delta) + Laplace(0, \lambda),
\end{equation}
where $\nabla\textit{\textbf{E}}$ denotes a general term for the gradient update matrix, in our case either $\nabla\textit{\textbf{E}}_{P_u}$ or $\nabla\textit{\textbf{E}}_{V_u}$, and $\delta$ denotes the clipping scale applied to the local gradients; 0 and $\lambda$ respectively denotes the location and the scale parameter of the Laplace distribution used to drawn a Laplace noise. The scale parameter $\lambda$ controls the strength of the Laplace noise such that larger $\lambda$ brings a larger privacy budget but a lower utility of the perturbed gradients. The function $clip(\nabla\textit{\textbf{E}}, \delta)$ is used to limit the value of the gradients in the scale of $\delta$. It is used to help avoid potential gradient explosion and to control the degree of sensitivity of the local gradients (more details in \Cref{subsec:privacy analysis}). After performing the clipping and the randomization operations, each client uploads its privatized local gradients -- $\nabla\widetilde{\textit{\textbf{E}}}_{V_u}$ and $\nabla\widetilde{\textit{\textbf{E}}}_{P_u}$ -- to the server where all the local gradients are then aggregated by averaging.
\begin{equation}
\label{eq:v_grad_aggre}
    \nabla\textit{\textbf{E}}_V = \frac{1}{|U|}\sum_{u\in U}\nabla\widetilde{\textit{\textbf{E}}}_{V_u}
\end{equation}
\begin{equation}
\label{eq:p_grad_aggre}
     \nabla\textit{\textbf{E}}_P = \frac{1}{|U|}\sum_{u\in U}\nabla\widetilde{\textit{\textbf{E}}}_{P_u}
\end{equation}
The aggregated gradients, $\nabla\textit{\textbf{E}}_V$ and $\nabla\textit{\textbf{E}}_P$, are then used to update the client-agnostic item and attribute matrices $\textit{\textbf{E}}_V$ and $\textit{\textbf{E}}_P$ stored on the server:
\begin{equation}
\label{item_matrix_update}
    \textit{\textbf{E}}_V = \textit{\textbf{E}}_V - \eta_2 \cdot \nabla\textit{\textbf{E}}_V
\end{equation}
\begin{equation}
\label{attr_matrix_update}
    \textit{\textbf{E}}_P = \textit{\textbf{E}}_P - \eta_3 \cdot \nabla\textit{\textbf{E}}_P,
\end{equation}
where $\eta_2$ and $\eta_3$ respectively denote the learning rate used for optimizing the item and the attribute matrix. The updated client-agnostic item and attribute matrices $\textit{\textbf{E}}_V$ and $\textit{\textbf{E}}_P$ are then distributed back to each client for further tuning. 

Once the system acquires an initial estimation of each user's historical interests, it enters the personalized \textit{interactive preference elicitation} stage to gain a deeper and more nuanced understanding of the user's current needs and preferences. To accomplish this, many existing CRSs \cite{crm, ear, scpr, UNICORN, FPAN, CRIF} employ a reinforcement learning based policy agent to customize the system's interactions (e.g. the attributes that the system inquires about) in order to deliver a personalized recommendation experience. This policy agent is typically trained through a trial-and-error process, in which it directly interacts with different users to learn \textit{an optimal interaction strategy} that maximizes cumulative rewards. These rewards are derived based on users' feedback to the current interaction strategy and are used as the numerical reflections to evaluate and train the current interaction strategy. In order to obtain these rewards, the majority of existing CRSs adopt the widely-accepted paradigm which requires users to share their raw feedback\footnote{A sample of such user feedback would be: \\ System: are you looking for more \textit{country} music? \\ User: Not really, I used to like \textit{country} music but now I am more into \textit{jazz}.} with the server. However, this approach introduces two challenges. First, storing human-interpretable conversational feedback, which often contain explicit and nuanced preferences of users, in a central repository dramatically increases the risk of unauthorized access and data breaches, which could lead to unwanted exposure of users' private information or even identity theft \cite{lam2006you}. Second, since each user has distinct needs and preferences, a uniform interaction strategy can only provide sub-optimal recommendation experience for each user, comparing to the oracle optimal interaction strategy \cite{jin2022federated}. 
\subsection{Decentralized Interactive Preference Elicitation}
\begin{figure}
  \includegraphics[width=.51\textwidth]{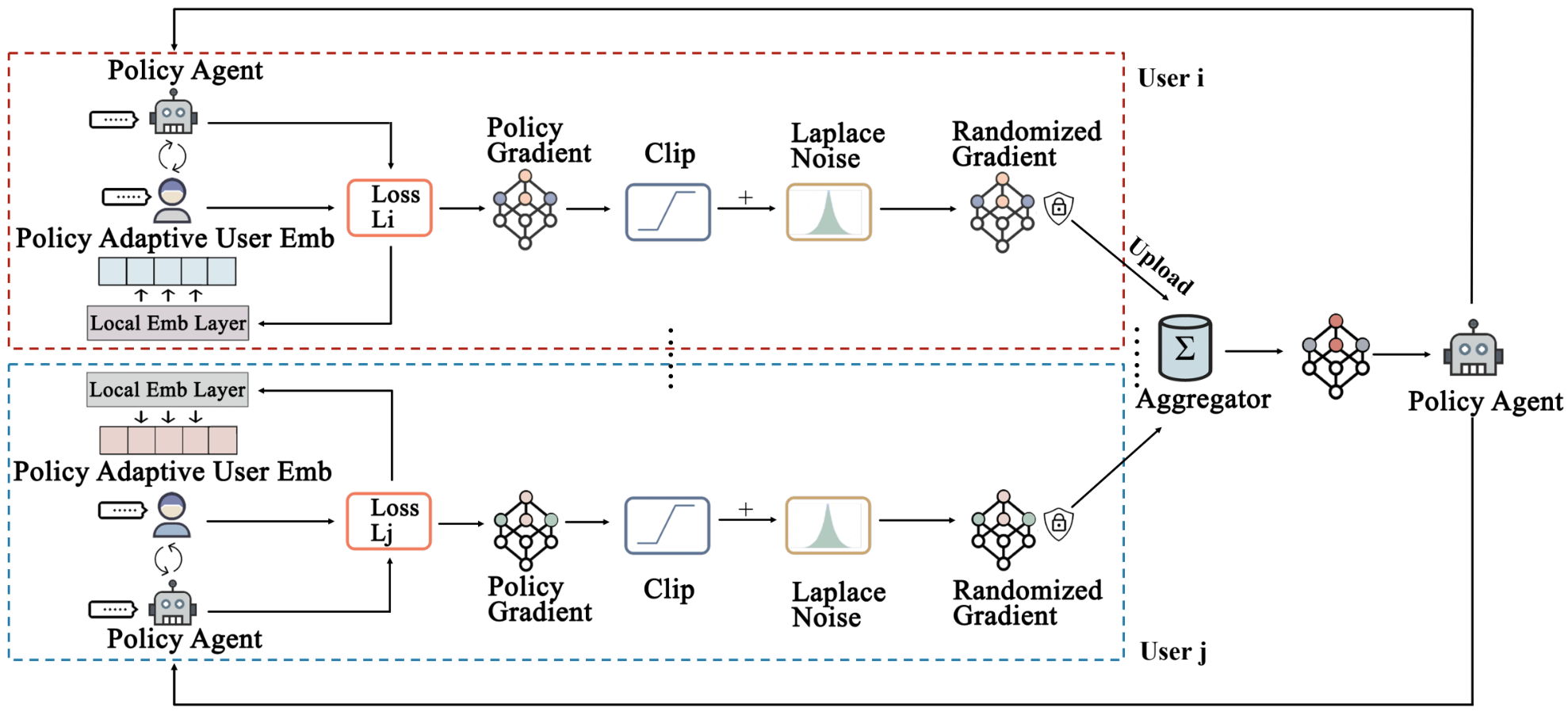}
  \caption{The overall flow of the decentralized interactive preference elicitation stage}
  \label{fig:model2}
\end{figure}

To address these two challenges, we propose a personalized federated reinforcement learning paradigm that privatizes the training process of the policy agent, while accommodating environment heterogeneity caused by each user having unique preferences. In our proposed paradigm, illustrated in \Cref{fig:model2}, all interactions between the policy agent and the user are confined to each user's own device(s), while the policy agent is collaboratively updated using aggregated contributions from various users. To address the environment heterogeneity that hinders a uniform policy from delivering optimal recommendation experiences to every user, we employ a learnable local user embedding projection layer to locally finetune the policy agent to adapt to each user's distinct profile. This learnable local user embedding layer is directly tuned at each user's device(s) and never shared with the server. Such approach enables the system to adapt the uniform policy to each user's specific needs and preferences, while providing an additional layer of privacy protection.  

To start, the server initializes the policy agent that will be trained collaboratively. To demonstrate the efficacy of the added user-local embedding layer, we choose a simple multi-layer perceptron neural network to serve as the policy agent that determines how the system interacts with the user. Specifically, the policy agent  is composed of two fully-connected layers that map the input state vector, $\textbf{s}_u$, into an action, $a$, in the action space, $A$. Following \cite{ear}, we define $A$ to be $\{a_{rec} \cup \{a_{ask}(p)|p \in P\}\}$. Formally, the policy agent is defined as follows:
\begin{equation}
\label{action_eq}
    a = \argmax \left[Softmax( Relu\left( \textbf{W}^{(2)} Relu\left(\textbf{W} ^{(1)} \mathbf{s}_u + \mathbf{b}^{(1)} \right) + \mathbf{b}^{(2)} \right) \right], 
\end{equation}
where $\textbf{W}$ and $\textbf{b}$ are the weights and biases respectively for the input and hidden layers; and $\textbf{s}_u$ denotes the user-specific input state vector:
\begin{equation}
\label{u_state}
    \textbf{s}_u = \textbf{s}_{emb} \oplus \textbf{s}_{hist},
\end{equation}
defined as the concatenation of (i) the output of the learnable local user embedding layer and (ii) the conversation history between the user and the system. The local user embedding projection layer 
\begin{equation}
\label{u_emb_layer}
    \textbf{s}_{emb} = Tanh\left(\textbf{W} ^{(u)} \textbf{e}_u + \textbf{b}_u \right)
\end{equation}
takes in the initial user embedding, $\textbf{e}_u$ (see \Cref{stage1}), as the input and outputs an policy-adaptive embedding, $\textbf{s}_{emb}$, that better suits the current interaction strategy. The conversation history, $\textbf{s}_{hist}$, is represented by a one-hot vector of dimensions equivalent to the total number of attribute, $|P|$, where the $i$-th dimension denotes whether the user has expressed preference on the attribute $p_i$.

Once the policy agent has been initialized on the server, its parameters $\theta$ are then shared with each client, signaling the start of a federated epoch. Each client interacts \textit{locally} with the client-agnostic policy agent to collect $N$ sequences of interactions between the client and the system, in the form of state-action-reward tuples. Each sequence is known as a sampled trajectory, defined as $\tau = [{(\textbf{s}_{u_1}, a_1, r_{u_1}), ..., (\textbf{s}_{u_t}, a_t, r_{u_t})}]$. In a sampled trajectory, $\textbf{s}_{u_i}$ denotes the user-specific state vector at the $i$-th time -- as specified in \Cref{u_state}; $a_i$ denotes the system's action choose by the policy agent with $\textbf{s}_{u_i}$ as the input -- as specified in \Cref{action_eq}; and $r_i$ denotes the numerical reflection (more details in \Cref{exp_setup}) of the user's feedback on the appropriateness of the chosen action, $a_i$. Based on the sampled trajectories, the policy agent is optimized locally with the objective to maximize the overall expected returns, defined as:
\begin{equation*}
    J(\theta)_u = \frac{1}{N} \sum_{i=1}^N \sum_t \gamma^t r(s_{i,t}, a_{i,t}),
\end{equation*}
where $\theta$ denotes the parameters of the policy agent shared from the server; $\gamma^t$ denotes the discount parameter emphasizing conversation efficiency. Following the REINFORCE algorithm \cite{sutton1999policy}, the local policy gradient of the learning objective can be defined as: 
\begin{equation}
\label{eq:reinforce}
    \nabla J(\theta)_u = \frac{1}{N} \sum_{i=1}^N \left[ \left( \sum_t \nabla_\theta log \pi_\theta(a_{i,t}|s_{i,t}) \right) \left( \sum_t r(s_{i,t}, a_{i,t})  \right) \right]
\end{equation}
Before sending local policy gradient updates, $\nabla J(\theta)$, back to the server, we apply the randomization algorithm $M$, introduced in \Cref{stage1_lap}, to perturb $\nabla J(\theta)$ to guarantee local differential privacy, for an enhanced user privacy protection:
\begin{equation}
\label{stage2_lap}
    \nabla \widetilde{J}_(\theta)_u = clip\left(\nabla J(\theta), \delta\right) + Laplace\left(0, \lambda \right),
\end{equation}
where $\delta$ denotes the clipping scale applied to the local gradients; and 0 and $\lambda$ respectively denotes the location and the scale parameter of the Laplace distribution used to drawn a Laplace noise. Once the clients have computed and uploaded their local policy gradients, the server aggregates all the collected local policy gradients by averaging:
\begin{equation*}
    \nabla J(\theta) = \frac{1}{|U|}\sum_{u \in U} \nabla \widetilde{J}_(\theta)_u
\end{equation*}
We use the averaged gradients to update the current policy agent:
\begin{equation*}
    \theta = \theta - \alpha \cdot \nabla J(\theta),
\end{equation*}
where $\theta$ denotes the parameters of the policy agent; $\alpha$ denotes the learning rate. It is important to note that, the parameters -- $\textbf{W}^{(u)}$ and $\textbf{b}_u$ -- of each user's local user embedding layer, specified in \Cref{u_emb_layer}, are never shared with the server and are only updated locally based on the client's sampled trajectories. This approach not only enhances personalization of the policy agent but also provides an additional layer of user privacy protection.

\subsection{Privacy Analysis}
\label{subsec:privacy analysis}
In this section, we formally prove the privacy guarantees of the proposed framework. First, we show that the randomized algorithm $M$ -- used in \Cref{stage1_lap} and \Cref{stage2_lap} -- that injects Laplace noise to perturb the local gradients guarantees $\epsilon$-local differential privacy. Let $\textbf{x}$ and $\textbf{y}$  be any neighboring local gradient updates of dimension $k$, uploaded by two different clients. Let $p_{\textbf{x}}(z)$ and $p_{\textbf{y}}(z)$ be the probability density functions of $M(\textbf{x})$ and $M(\textbf{y})$ evaluated at an arbitrary point $z\in\mathbb{R}^k$. We show that the ratio of $p_{\textbf{x}}(z)$ to $p_{\textbf{y}}(z)$ is bounded above by $e^\epsilon$:
\begin{align*}
    \frac{P_\textbf{x}(z)}{P_\textbf{y}(z)} &= \frac{\prod_{i=1}^k exp(-\frac{\epsilon|\textbf{x}_i - z_i|}{\triangle})}{\prod_{i=1}^k exp(-\frac{\epsilon|\textbf{y}_i - z_i|}{\triangle})} \\
    &= \prod_{i=1}^k exp({-\frac{\epsilon(|\textbf{x}_i - z_i| - |\textbf{y}_i - z_i|)}{\triangle}}) \\
    &\leq \prod_{i=1}^k exp({-\frac{\epsilon|\textbf{x}_i - \textbf{y}_i|}{\triangle}}) \\
    &= exp (\frac{\epsilon\sum_{i=1}^k|\textbf{x}_i - \textbf{y}_i|}{\triangle}) \\ 
    &= exp (\frac{\epsilon||\textbf{x} - \textbf{y}||_1}{\triangle}) \\
    &\leq exp(\epsilon), 
\end{align*}
This proves that the randomized algorithm $M$ guarantees local differential privacy, such that an \textbf{adversary cannot confidently distinguish} an individual client's local gradient updates from the local gradient updates of another user \cite{xiong2016randomized}. The $\triangle$ denotes the $l_1$ sensitivity, between any two neighboring local gradient updates $\textbf{x}$ and $\textbf{y}$, defined as: 
\begin{equation*}
    \triangle = max_{\textbf{x}, \textbf{y}}||\textbf{x} - \textbf{y}||_1.
\end{equation*}
The $\epsilon$ denotes the privacy budget which for the applied randomized algorithm $M$, $\epsilon$, and it is defined as: 
\begin{equation}
\label{eq:eps}
    \epsilon = \frac{\triangle}{\lambda},
\end{equation}
where $\lambda$ is the scale parameter used to draw the Laplace noise. Note, since a clipping function is applied in the randomized algorithm $M$ -- \Cref{stage1_lap} and \Cref{stage2_lap}, the $l_1$ sensitivity, $\triangle$, is upper bounded by $2\delta$, with $\delta$ being the clipping scale. Thus, from \Cref{eq:eps}, we can observe that the privacy budget, $\epsilon$, is upper bounded by $\frac{2\delta}{\lambda}$.     

\subsection{Communication Cost}
\label{subsec:communication}
For the \textit{historical user interests estimation} stage, at the start of each federated epoch, each client receives an $M \times F$ real-valued item matrix $\textit{\textbf{E}}_V$ and an $N \times F$ real-valued attribute matrix $\textit{\textbf{E}}_P$, where $M$ and $N$ respectively denotes the number of items, and the number of attributes in the dataset, and $F$ denotes the dimensionality of item and attribute latent embedding (e.g., 64). Depending on the size of the dataset, the size of this message can vary greatly. Consider the two datasets in the experiments (see Section~\ref{sec:experiments}). For the LastFM dataset, the combined size of $\textit{\textbf{E}}_V$ and $\textit{\textbf{E}}_P$ is about 0.9 MB; however, for the Yelp dataset, the combined size adds up to 8.5 MB. Once the local gradient updates have been computed, each clients uploads the item gradient matrix $\nabla\widetilde{\textit{\textbf{E}}}_{V_u}$ and the attribute gradient matrix $\nabla\widetilde{\textit{\textbf{E}}}_{V_u}$ to the server. Since $\nabla\widetilde{\textit{\textbf{E}}}_{V_u}$ and $\nabla\widetilde{\textit{\textbf{E}}}_{V_u}$ share the same size as $\textit{\textbf{E}}_V$ and $\textit{\textbf{E}}_P$, the total size of each federated epoch for the LastFM and the Yelp dataset respectively sums up to 1.8 MB and 17 MB. Note, while the communication cost could scale up as the size of the dataset increases, there are many compression techniques \cite{rw5_no_crs, compress2, compress3, compress4} that can be easily integrated to mitigate this issue; we leave the integration of communication cost compression for future work. For the \textit{historical user interests estimation} stage, during each federated epoch, each client receives a copy of the parameters, $\theta$, of the policy agent, and uploads the locally computed policy gradient $\nabla \widetilde{J}_(\theta)_u$ to the server. The communication cost sums up to $2|\theta|$, which yields to 0.03 MB for the LastFM dataset and 0.17 MB for the Yelp dataset.

\begin{table}
  \begin{center}
  \setlength{\tabcolsep}{5pt}
    \caption{Dataset statistics}
    \label{tab:table1}
    \begin{tabular}{l r r r r}
      \hline
      Dataset & \#users & \#items & \#interactions & \#attributes \\
      \hline  
      LastFM & 1,801 & 7,432 & 76,693 & 33 \\
      Yelp & 27,675 & 70,311 & 1,368,606 & 590 \\
      \hline
    \end{tabular}
  \end{center}
\end{table}

\begin{table*}[h]
  \setlength{\tabcolsep}{6pt}
  \label{overall_perf}
  \caption{Overall Recommendation Performance Comparison, with SR@15 $\uparrow$ and AT $\downarrow$used as evaluation metrics. Boldface used to highlight the best performance, while underline highlights the runner-up.}
  \begin{center}
    \begin{tabular}{ c c c c c c c c c c c } 
        \hline
         &  & AbsGreedy & MaxEnt & CRM & EAR & SCPR & UNICORN & FPAN & CRIF & FedCRS\\
        \hline
        \multirow{2}{4em}{LastFM} & SR@15 & 0.222 & 0.283 & 0.325 & 0.429 & 0.465 & 0.535 & 0.631 & \textbf{0.866} & \underline{0.831} \\ 
        & AT & 13.48 & 13.91 & 13.75 & 12.88 & 12.86 & 11.82 & 10.16 & \textbf{9.17} & \underline{9.54} \\
        \hline
        \multirow{2}{4em}{Yelp} & SR@15 & 0.264 & 0.921 & 0.923 & 0.957 & 0.973 & \underline{0.984} & 0.979 & \textbf{0.986} & 0.974 \\
        & AT & 12.57 & 6.59 & 6.25 & 5.74 & 5.67 & 5.33 & 5.08 & \textbf{4.35} & \underline{4.92} \\ 
        \hline
    \end{tabular}
  \end{center}
\end{table*}

\section{Experiments}
\label{sec:experiments}
We conduct experiments to evaluate the recommendation performance and the privacy budget trade-off of the proposed FedCRS approach. Specifically, we focus on answering three key research questions: \textbf{RQ1}. How does recommendation performance of the proposed FedCRS framework compare to existing non-private CRS approaches? \textbf{RQ2}. How does the estimated user historical interests and the personalized policy agent respectively contribute to the overall recommendation performance? \textbf{RQ3}. How does the privacy budget of the randomized algorithm affect the \textit{recommendation performance} of FedCRS? Note, during the page limit, we include the training details in our supplementary material.

\subsection{Experiment Setup}
\label{exp_setup}
\noindent\textbf{Dataset.} We evaluate the proposed framework on two benchmark datasets -- Yelp \cite{crm} for business recommendation and LastFM \cite{ear} for music artist recommendation -- that are widely adopted \cite{crm, ear, scpr, CRIF, UNICORN} to evaluate conversational recommendation systems. We follow the common recommendation evaluation setting \cite{he2017neural, DBLP:journals/corr/abs-1205-2618} to prune users with fewer than 10 item interactions and split user-item interactions in a ratio of 7:2: 1 for training, validation, and testing. For the Yelp dataset, we perform the same hierarchical attribute prepossessing as in previous works \cite{ear, scpr, FPAN}. The two datasets are summarized in \Cref{tab:table1}.

\noindent\textbf{Baselines.} To validate the proposed framework, we compare its performance with the following baseline conversational recommendation approaches. Note, all baseline approaches are both trained and deployed in a centralized manner that fails to meet any of the defined privacy protection guides defined in \Cref{sec:intro}. 
\squishlist
  \item \textbf{Absolute Greedy (AbsGreedy)} solely recommends items in every interaction. Once the recommendation is rejected, it excludes these items from the candidate itemset. 
  \item \textbf{Max Entropy (MaxEnt)} adopts a rule-based policy agent that always selects the attribute with the highest entropy to ask.
  \item \textbf{CRM \cite{crm}} uses a belief tracker to record a user's conveyed preference, and a policy agent to decide how to interact.
  \item \textbf{EAR \cite{ear}} builds a predictive model to estimate a user's historical preference and trains a policy agent to determine hwo to interact. In addition, it also considers the feedback from the user to further fine-tune the learned predictive model.
  \item \textbf{SCPR \cite{scpr}} leverages the concept of adjacent attributes to construct a knowledge graph to reduce attribute search space and uses Deep Q-Learning \cite{van2016deep} to learn a more efficient policy agent.
  \item \textbf{UNICORN \cite{UNICORN}} applies a dynamic-weighted-graph based reinforcement learning approach to integrate the conversation and the recommendation components and deploys an weighted-entropy action selection strategy to infer the candidate actions.
  \item \textbf{FPAN \cite{FPAN}} utilizes a user-item attribute graph to enhance the embeddings learned by the predictive model, then applies a gating mechanism to aggregate the user feedback to constantly refine the learned user embeddings. 
  \item \textbf{CRIF \cite{CRIF}} is the current state-of-the-art CRS. It propose a novel inference module that explicitly infer a user's implicit preference conveyed during the conversation, and adopts inverse reinforcement learning to learn a flexible policy agent which makes the dialogue more coherent.
\squishend

\noindent\textbf{Evaluation Metrics.}
To evaluate the recommendation performance of a CRS, we use two widely adopted \cite{crm, ear, scpr, CRIF, UNICORN, FPAN} metrics -- success rate at turn t (SR@t) and average turn (AT) of conversations. SR@t is the accumulative ratio of successful conversational recommendation by the turn t, and AT is the average number of turns for all sessions.

\noindent\textbf{Training Details.}
The training process is divided into an offline and an online stage. The offline stage is intended for training a predictive model that accurately estimates the historical interests of users, while the online training is intended for training a policy agent that effectively facilitates personalized interactive preference elicitation. For training the predictive model, we set dimensionality, $d$, of all the user, item, and attribute embeddings to 64. For the randomized algorithm $M$ in \Cref{stage1_lap}, the clipping scale, $\delta$, for both the item and the attribute matrices is set to 0.0025. And the scale parameter, $\lambda$, used to draw the Laplace noise is set to 0.01. The Dual-BPR objective function -- \Cref{eq:dual_loss_bpr} -- along with back-propagation are used to compute the gradient updates for the local user embedding, and the two client-agnostic item and attribute matrices. We set $\eta_1$ in \Cref{u_emb_update} to .01, $\eta_2$ in \Cref{item_matrix_update} to 1.5, and $\eta_3$ in \Cref{attr_matrix_update} to 2. Note that much greater learning rates are used for updating the item and the attribute matrices since their gradients have been clipped into a much smaller range -- see \Cref{stage1_lap}. For the reinforcement learning based policy agent specified in \Cref{action_eq}, we design it to take in output of the user local embedding layer concatenated with a one-hot $\textbf{s}_{pref}$ vector as specified in \Cref{u_state}. The user embedding layer, specified in \Cref{u_emb_layer}, is simply a single-layer perceptron with both the input and output size equivalent to $|\textit{\textbf{e}}_u|$. The policy agent is composed of two fully-connected layers with parameter sizes of $|P|+|\textit{\textbf{e}}_u|$ for the input layer, and 64 for the hidden layer. To train it, we use \Cref{eq:reinforce} to compute the local policy gradients with three types of rewards defined as: (1) $r_{suc} = 1$, a strongly positive reward for successful recommendation; (2) $r_{ask} = 0.25$, a smaller positive reward when the user confirms the asked attribute(s); and (3) $r_{quit} = -1$ a strongly negative reward if the system fail to reach successful recommendation within $t$ interactions. To privatize the computer local policy gradients, we set $\delta$ and $\lambda$ in \Cref{stage2_lap} respectively to 0.0025 and 0.01, resulting in a privacy budget of 0.5, which, as shown in \cite{minto2021stronger_rec21, qi2020privacy_news, gao_sigir_2020dplcf}, provides a strong privacy protection.

\noindent\textbf{User Simulator for Conversational Recommendation.}
Since conversational recommendation is a dynamic process, we follow \cite{crm, ear, scpr, CRIF, UNICORN, FPAN} to create a user simulator to enable training and evaluation of a CRS. We simulate a conversation session for each observed interaction between users and items. Specifically, given an observed user–item interaction $(u,v)$, we treat the v as the ground truth item and consider its attributes $P_v$ as the oracle set of attributes preferred by the user in this conversation session. At the beginning, we randomly choose an attribute from the oracle set as the user’s initial request to the CRS (e.g., Hi, can you find a \textit{country} music?). Then the conversation session goes in the loop of the ``model asks – user responses'' process as introduced in \Cref{prelim_crs}. The session ends when either the CRS has recommended the groundtruth item $v$ or a pre-defined number of allowed interactions has been reached.

\subsection{Recommendation Performance (RQ1)}
\label{subsection:RQ1}

\begin{figure}
  \includegraphics[width=.49\textwidth]{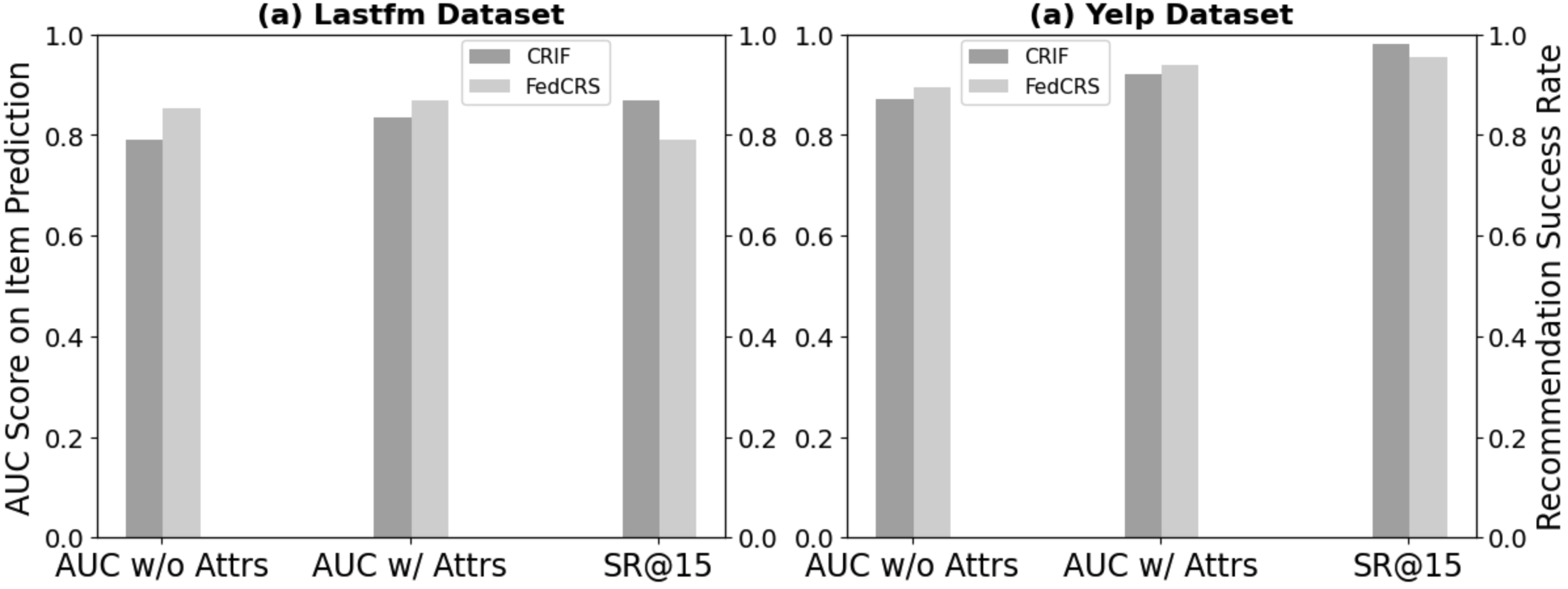}
  \caption{Effectiveness of learned embeddings and adopted policy agent in FedCRS and CRIF}
  \label{fig:performance_comp}
\end{figure}

As shown in \Cref{overall_perf}, we can observe that the proposed FedCRS is able to outperform all the baseline CRSs expect for CRIF on the LastFM dataset. And, although both UNICORN and CRIF achieve a higher SR@15 than FedCRS on the Yelp dataset, the gap amongst them is relatively small. It is also worth noting that all CRSs achieve better recommendation performance (both in SR@15 and AT) on the Yelp dataset. This is because Yelp adopts an enumerated question setting, which allows a CRS to collect user preferences on multiple attributes simultaneously. Since the primary intention of FedCRS is to preserve user privacy in conversational recommendations, it is perhaps surprising that it demonstrates a competitive performance even when compared to CRIF. We conduct additional experiments to understand the factors contributing to this phenomenon. In \Cref{fig:performance_comp}, we compare both: (1) the effectiveness of the embeddings learned respectively by FedCRS and CRIF; and (2) the effectiveness of policy agents adopted respectively by FedCRS and CRIF. 

To compare the effectiveness of the learned embeddings, we report their respective AUC scores on item prediction both with and without considering each item's associated attributes. The former task measures the effectiveness of the learned user and item embeddings, while the latter emphasizes on the learned attribute embeddings. As observed, FedCRS achieves higher AUC scores on both tasks than CRIF. This is primarily due to the \textit{unweighted} aggregation of local matrices gradients, specified in \Cref{eq:v_grad_aggre} and \Cref{eq:p_grad_aggre}. With an unweighted aggregation, every user is deemed equally important to the objective function. This approach helps the objective function to simultaneously optimize the embeddings of all users, rather than disproportionately focusing on more active users and neglecting those with fewer item interactions.\footnote{Note this is a frequently encountered problem that hinders the overall recommendation performance of recommender systems trained in a centralized manner \cite{10.1145/3038912.3052713, lin2022quantifying}} 

To evaluate the effectiveness of the learned policy agents, we employ \textit{CRIF's learned embeddings} in conjunction with the FedCRS's policy agent and report its SR@15. In essence, we maintain the \textit{same} embeddings for both FedCRS and CRIF and exclusively compare the effectiveness of their respective policy agents. As observed in \Cref{fig:performance_comp}, while using the same embeddings, the policy agent proposed in CRIF is able to slightly outperform FedCRS on both the LastFm and the Yelp dataset. This is less surprising since the primary objective of this work is to propose a decentralized framework for the privacy-preserving training and deployment of CRSs. We leave the integration of more sophisticated policy agents for future work. 

\subsection{Ablation Studies (RQ2)}
\label{subsection:RQ2}

We next conduct ablation studies to further assess how each of the major design components -- \textit{Decentralized Historical Interests Estimation} and \textit{Decentralized Interactive Preference Elicitation} -- contributes to the overall recommendation performance. 

For the \textit{Decentralized Historical Interests Estimation} component, we measure its contribution by replacing the learned embeddings with (1) randomly initialized embeddings; and (2) CRIF's learned embeddings, which have been shown to be less effective than FedCRS's in \Cref{subsection:RQ1}. As observed in \Cref{tab:ab_study}, replacing the learned embeddings in FedCRS with randomly initialized embeddings causes dramatic drop in recommendation performance (both in SR@15 and AT). In addition, replacing FedCRS's embeddings with CRIF's embedding also incurs a decrease in recommendation performance. This greatly demonstrates the importance of the embeddings learned by the proposed \textit{Decentralized Historical Interests Estimation} component. For the \textit{Decentralized Interactive Preference Elicitation} component, the main novelty lies in the added local user embedding projection layer, as specified in \Cref{u_emb_layer}. Since previous works \cite{jin2022federated, federated_survey3} have shown that a uniform reinforcement learning based policy agent can only achieve sub-optimal performance due to environment heterogeneity. We apply a \textit{learnable} local user embedding layer to transform the input of the policy agent to better adapt to the existing policy. As observed in \Cref{tab:ab_study}, removing the learnable local user embedding layer causes significant decrease in recommendation performance (both in SR@15 and AT). This demonstrates the importance of the local user embedding layer to the \textit{Decentralized Interactive Preference Elicitation} component.

\subsection{The Privacy Budget Trade-off (RQ3)}
\label{subsec:tradeoff}
As introduced in \Cref{subsect:LDP}, the privacy budget, $\epsilon$, controls the trade-off between the degree of user privacy preserved by the randomized algorithm, $M$, and the utility of the perturbed local gradients. Since, $\epsilon$, is upper bounded by $\frac{2\delta}{\lambda}$ (see \Cref{subsec:privacy analysis}), increasing $\lambda$ helps $M$ to achieve a \textit{smaller} privacy budget which indicates \textit{better} privacy protection; however, increasing $\lambda$ also causes stronger injected noises which could hurt the accuracy of aggregated gradients, resulting in a less effective predictive model or policy agent. In this section, we thoroughly exam the privacy budget and recommendation performance trade-off in FedCRS. As shown in \Cref{subsection:RQ1}, the recommendation performance of FedCRS is closely related to the quality of embeddings learned by the predictive model during the \textit{historical user interest estimation} stage and the effectiveness of the policy agent trained during the \textit{interactive preference elicitation} stage. Therefore, to study the relationship between privacy budget and recommendation performance, we respectively investigate how different values of privacy budget respectively affect the effectiveness of the trained predictive model and policy agent.

\begin{table}
  \label{tab:ab_study}
  \caption{Ablation Studies with SR@15 $\uparrow$ and AT $\downarrow$ as metrics.}
  \begin{center}
    \begin{tabular}{l c c c c} 
      \hline
      \multicolumn{1}{c}{ } & \multicolumn{2}{c}{Lastfm} & \multicolumn{2}{c}{Yelp} \\
      \multicolumn{1}{c}{ } & SR@15 & AT & SR@15 & AT \\
      \hline
      w/ Random Embeddings & 0.251 & 13.96 & 0.327 & 12.05 \\
      
      w/ CRIF Embeddings & 0.784 & 10.14 & 0.951 & 5.73 \\
      
      w/o Local Embedding Layer & 0.722 & 11.07 & 0.919 & 6.14 \\
      \hline
      \textbf{FedCRS} & \textbf{0.831} & \textbf{9.54} & \textbf{0.974} & \textbf{4.92} \\
      \hline
    \end{tabular}
  \end{center}
\end{table}

\begin{figure}
  \includegraphics[width=.48\textwidth]{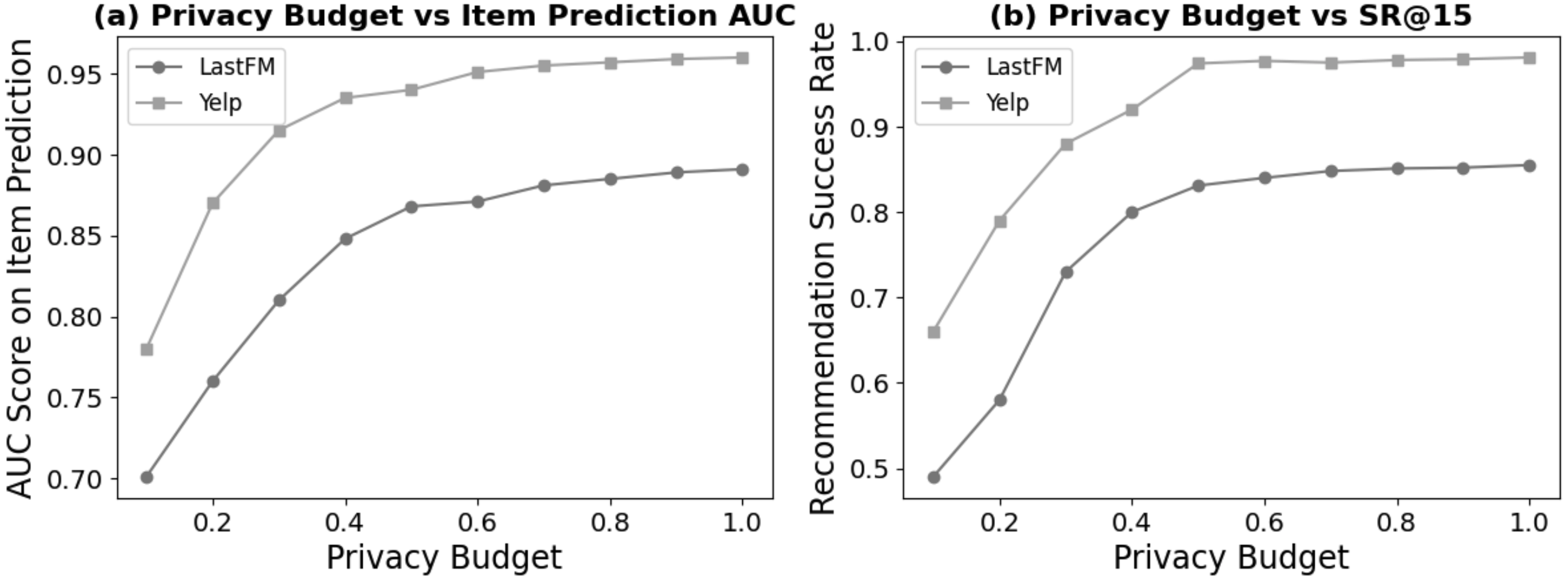}
  \caption{Effects of different privacy budgets on effectiveness of the train predictive model or policy agent}
  \label{fig:privacy_budget}
\end{figure}

We use AUC scores of the learned embeddings as a metric to measure the effectiveness of the trained predictive model. As observed in \Cref{fig:privacy_budget}(a), large privacy budgets enable FedCRS to train a more effective predictive model. A similar phenomenon is also observed in \Cref{fig:privacy_budget}(b), where the recommendation success rate of the trained policy agent increases with the privacy budget, indicating larger privacy budgets facilitate the learning of a more personalized interaction strategy. These observations align with our intuition that although smaller privacy budgets provide stronger privacy protection, the increased noise level (large $\lambda$ in $M$) also greatly limits the utility of the uploaded local (policy) gradients. It is worth noting that both \Cref{fig:privacy_budget}(a) exhibit a diminishing increase in slope for privacy budgets greater than 0.5. Thus, we select 0.5 as our privacy budget, as it achieves an optimal compromise between user privacy preservation and recommendation performance.

\section{Related Work}
\subsection{Conversational Recommender System}
Conversational recommender systems have emerged as an increasingly popular tool to significantly personalize a user's recommendation experience \cite{FPAN, UNICORN, CRIF}. While early works on CRSs primarily rely on template conversation components \cite{jiang2014choice, graus2015improving}, recent works have explored the possibilities of more effective preference elicitation methods and conversational strategies with the help of reinforcement learning. For example, CRM \cite{crm} proposes a deep reinforcement-based conversation component that builds a personalized preference elicitation process. Inspired by CRM, EAR \cite{ear} proposes a novel three-staged framework that further strengthens the interaction between the recommender component and the conversation component. While sharing the same recommender component, SCPR \cite{scpr} integrates knowledge graphs to improve the reasoning ability of the conversation component. Further utilizing the informativeness of knowledge graphs, UNICORN \cite{UNICORN} proposes a dynamic weighted graph based reinforcement approach. Furthermore, CRIF \cite{CRIF} introduces the adaption of inverse reinforcement learning to learn a more flexible conversation policy, making the generated dialogue more coherent.

\label{subsection:RQ3}

\begin{table}
  \label{tab:ab_study}
  \caption{Warm vs. Cold Item Performance Comparison with AUC $\uparrow$ and F1 Score $\downarrow$ as metrics.}
  \begin{center}
    \begin{tabular}{l c c c c} 
      \hline
      \multicolumn{1}{c}{ } & \multicolumn{2}{c}{ML-1M} & \multicolumn{2}{c}{Amazon Kindles} \\
      \multicolumn{1}{c}{ } & AUC & F1 Score & AUC & F1 Score \\
      \hline
      Ensemble Avg - warm & 0.8171 & 0.7675 & 0.8835 & 0.8103 \\
      DCN - warm & 0.7946 & 0.7591 & 0.8520 & 0.7721 \\
      \hline
      Ensemble Avg - cold & 0.7142 & 0.6297 & 0.8756 & 0.8003 \\
      DCN - cold & 0.6566 & 0.5869 & 0.8279 & 0.7472 \\
      \hline
    \end{tabular}
  \end{center}
\end{table}

Although CRSs has shown to provide promising recommendation performance \cite{scpr, CRIF, FPAN}, most of the existing CRSs centrally collect each user's complete \textit{item-interaction history} and \textit{conversation history} for training and inference. As shown in previous works \cite{massa2009trust, cormode2018privacy, jeckmans2013privacy}, such a centralized training paradigm poses significant privacy concerns, especially when the uploaded data contains thoroughly communicated - human interpretable - user preferences. In this work, we propose a federated conversational recommendation framework that address the privacy concerns while achieving strong recommendation performance. 

\subsection{Privacy-Preserving Recommendation}
Recommender systems help users navigate information by linking them to relevant items but also raise substantial privacy concerns \cite{cormode2018privacy}. Studies \cite{qi2020privacy_news, gao_sigir_2020dplcf} indicate that even seemingly innocuous data, like movie ratings, can reveal sensitive user information, such as health status or political views. These findings have spurred the advancement of privacy-preserving recommendation. Research in this area typically divides based on trust in the recommender. 

When the recommender is assumed to be \textit{trusted}, attacks aim to infer private information of the user from the released recommendation model (e.g., user and item embeddings) or the recommendation results \cite{gao_sigir_2020dplcf}. To protect the released model, \cite{chen2018privacy} proposes to split the learned user-item embeddings into local and global segments. For protecting the recommendation results, \cite{riboni2012private, berlioz2015applying} propose different means to perturb the results to satisfy differential privacy. While these works show promising results, assuming recommenders' trustworthiness is often unrealistic \cite{gao_sigir_2020dplcf}. Therefore, many works aim to preserve user privacy when the recommender is assumed to be \textit{distrusted}. Broadly, existing solutions can be divided into two categories - private data collection \cite{polat2005svd, li2017differentially, gao_sigir_2020dplcf} and distributed learning \cite{rw1_no_ldp, rw2_no_ldp, rw3_no_ldp, rw4_no_crs}. Data collection methods privatize the raw interaction data of the user before sharing it with the recommender. Early works in this direction include injection of Gaussian noise in the reported interaction data  \cite{gemulla2011large, li2017differentially}. More recently, \cite{minto2021stronger_rec21} explores different local perturbations on data to guarantee local differential privacy before uploading to the recommender. Different from data collection approaches, distributed learning approaches distribute the training process of the model, such that user interaction data is always kept on the client's device(s) and only gradient updates are uploaded to the server. However, since gradients can also be exploited to infer private information about the user \cite{wu2021fedgnn}, \cite{qi2020privacy_news} proposes to inject noise into the local gradients to ensure that the uploaded gradients meets local differential privacy guarantees. 

While the aforementioned approaches provide solutions for preserving user privacy in a static recommendation setting, none of them can be directly applied to privatize the explicitly communicated user preferences in a conversational recommendation setting. In this work, we propose a novel distributed framework for training and deploying a conversational recommender system under the assumption that the server cannot be trusted.

\section{Conclusion}
In this work, we present, to the best of our knowledge, the first comprehensive study of user privacy concerns in CRSs and introduce a set of guidelines aimed at protecting user privacy within the conversational recommendation setting. To comply with these guidelines, we propose a novel federated conversational recommendation framework (FedCRS) that effectively privatizes both the \textit{historical user interests estimation} stage and the \textit{interactive preference elicitation} stage by enforcing user-level local differential privacy. Through extensive experiments, we demonstrate that the proposed FedCRS not only satisfies all user privacy protection guidelines, but also provides strong recommendation performance.

\section{Ethical Considerations}
Interestingly, ethical concerns are among the primary motivations behind this work. To our knowledge, this is the first study to systematically address user privacy issues in conversational recommender systems. While our proposed framework aims to guide practitioners and researchers in integrating user privacy protection into these systems, it's essential to recognize potential challenges and repercussions. One potential issue is that our framework might unintentionally offer a blueprint for malicious attackers. With deeper insight into the system’s design, attackers could devise increasingly sophisticated strategies that target its latent vulnerabilities. Furthermore, as outlined in Section 4.4, the enhanced privacy guarantee does introduce trade-offs concerning recommendation quality. Even though the impact of this trade-off is relatively minor, it could lead to perceived disparities for some users and items, causing potential fairness concerns. We leave how to more effectively alleviate the privacy-performance compromise for future work.

\bibliographystyle{ACM-Reference-Format}
\bibliography{ref}

\end{document}